\documentstyle[aps,prl,twocolumn,epsf]{revtex}

\begin{document}
\input{psfig.sty}
\draft
\twocolumn[\hsize\textwidth\columnwidth\hsize\csname
@twocolumnfalse\endcsname

\title{\bf  
Charge and Spin Order in $\rm \bf La_{2-x}Sr_xNiO_4$ (x=0.275 and 1/3)}
\author{S.-H.  Lee$^{1,2}$, 
S-W. Cheong$^3$, K. Yamada$^{4}$, 
and C.F. Majkrzak$^1$}
\address{$^{1}$NIST Center for Neutron Research, National
Institute of Standards and Technology, Gaithersburg, MD 20899}
\address{$^{2}$University of Maryland, College Park, Maryland 20742}
\address{$^3$Department of Physics and Astronomy, Rutgers University and Bell Laboratories, Lucent Technologies, Murray Hill,
NJ 07974}
\address{$^4$Kyoto University, Institute of Chemical Research, Uji 611-0011,
Kyoto, Japan}
\maketitle
\begin{abstract}
We report polarized and unpolarized neutron diffraction measurements on 
$\rm La_{2-x}Sr_xNiO_4$ (x=0.275 and 1/3).
The data for the spin ordered states are consistent 
with a collinear spin model in which spins with S=1 are in the NiO$_2$ plane
and are uniformly rotated from the charge and spin stripe direction.
The deviation angle is larger for x=1/3 than for x=0.275.
Furthermore, for x=1/3 the ordered spins 
reorient below 50 K, 
which suggests a further lock-in of the doped holes
and their magnetic interactions with the S=1 spins.

\end{abstract}


]

\newpage

Since the discovery of incommensurate (IC) magnetic fluctuations
in superconducting cuprates\cite{cheong91}, magnetism
in doped antiferromagnets has attracted much attention.
Doping holes into the antiferromagnetic (AFM) insulator $\rm La_2CuO_4$
rapidly weakens the static magnetic 
correlations and leads to a metallic, superconducting phase at 
moderate hole concentrations\cite{yamada98}. 
The discovery of static IC spin and charge peaks 
in $\rm La_{1.6-x}Nd_{0.4}Sr_xCuO_4$\cite{tranq95}
led to a model of charge stripes acting as
antiphase domain walls for the intervening 
AFM regions\cite{tranq95,emery99}.

$\rm La_2NiO_4$ is a Mott insulator with $T_N\simeq 650$ K and 
spins aligned with the shortest orthorhombic 
lattice direction\cite{aeppli88}.
Doping of divalent Sr$^{2+}$ ions into the trivalent La$^{3+}$ sites in
$\rm La_{2-x}Sr_xNiO_4$ (LSNO(x)) rapidly suppresses
the ordering but the compound remains insulating until $x\approx 1$
where it becomes metallic.
For $x>0.1$, IC charge and spin superlattice reflections
appear, and can be characterized by $Q_{spin}=(1\pm \epsilon,0,0)$ 
and $Q_{charge}=(2\epsilon,0,1)$ 
with $\epsilon \approx x$\cite{hayden92,chen93,tranq94,yoshizawa}.
For $x\approx 1/3$, there is a commensurability effect which favors
$\epsilon =1/3$\cite{yoshizawa}.

Considering that the static spin structure should be a basis
for understanding the nature of spin and charge stripes
in the doped nickelates and cuprates,
it is surprising that the local spin structure,
evolving upon doping, has not been studied.
In this letter we report polarized and unpolarized neutron diffraction 
measurements on LSNO(x=0.275 and x=1/3) 
and present a collinear spin model to explain our data
in which spins are rotated uniformly by an angle $\theta$ 
from the stripe direction.
For x=0.275, the angle $\theta$ is $27(7)^o$ below $T_N=155(5)$ K.
For x=1/3, $\theta=40(3)^o$ for 50 K $<T<T_N=200(10)$ K
and below 50 K 
another phase transition occurs involving 
reorientation of spins to $\theta=53(3)^o$.
We argue that the spin reorientation is due to a further localization
of the doped holes on the lattice, which is consistent with
recent resistivity measurements that showed a
delocalization of charge stripes by an electic field 
(although in the lowest temperature phase the charge stripes are more
robust.)\cite{tokura}
Our polarized neutron data also provide
unequivocal evidence that
upon cooling the charge order precedes the spin order in these
materials.  Charge orders at 200(10) K for x=0.275, 
and at 240(10) K for x=1/3.

The LSNO(x) crystals with x=0.275 and x=1/3 used in this work were
grown by the floating-zone method
at Kyoto University and at Bell Laboratories, respectively.
The crystal structure of both materials remains tetragonal with the I4/mmm
symmetry over the range of temperature,
10 K $< T <$ 320 K, with $a_t=3.82(1)$~\AA~ and $c=12.64(5)$~\AA~
at 11 K for x=0.275 and with $a_t=3.83(1)$~\AA~ and $c=12.69(5)$~\AA~
at 11 K for x=1/3. For convenience, however, we use
orthorhombic lattice units with $a_o=\sqrt{2}a_t$ which is rotated
by 45$^o$ with respect to the Ni-O bonds in the NiO$_2$ planes.

Neutron scattering measurements were performed
on the cold neutron triple axis spectrometer SPINS at NIST. 
The incident neutron energy was $E_i=5$ meV and higher order
contamination was eliminated by a cryostat-cooled Be filter before
the sample. The incident neutrons were polarized in one spin state
by a forward transmission polarizer. The spin state of the scattered
neutrons from the sample was then analyzed with a rear flipper
and polarizer combination. A collimator was placed right after
each polarizer to eliminate the other spin state.
The samples were mounted in such a way that the (h0l) reciprocal
plane becomes the scattering plane. A guide field
was applied vertically along the (010)-axis. 
Collimations were guide-40-40-open for the LSNO(x=0.275) measurements 
and guide-40-20-open for the LSNO(x=1/3) measurements.
The polarizing efficiency was 0.89(1) for the former and 0.90(1) for the latter.
In this geometry, the spin component perpendicular to the
wavevector transfer $\vec{Q}$ 
which is parallel with the guide field
together with the nuclear structural component,
contribute to non-spin-flip (NSF)
channel whereas the spin component normal to $\vec{Q}$
and to the guide field contributes to the spin-flip (SF) channel.
Since spins in the nickelates are coplanar in ab-plane\cite{aeppli88,hk0}, 
the NSF and SF neutron scattering cross sections 
for $Q=(h0l)$, $\sigma_{NSF} (Q)$ 
and $\sigma_{SF} (Q)$ respectively, can be written as\cite{moon}
\begin{eqnarray}
\sigma_{NSF} (Q)& = & \sigma_N (Q)+\sigma_M^b (Q)\nonumber \\
\sigma_{SF} (Q)& = & 
		\sigma_{M}^a (Q)
 \left(1-\frac{(ha^*)^2}{(ha^*)^2+(lc^*)^2}\right)
\end{eqnarray}
where $\sigma_{N}$ is the structural scattering cross section,
$\sigma_{M}^{a}$ and $\sigma_{M}^{b}$ are the a- and b-component of 
the magnetic scattering cross section, respectively.

\vspace{0.15in}
\noindent
\parbox[b]{3.4in}{
\psfig{file=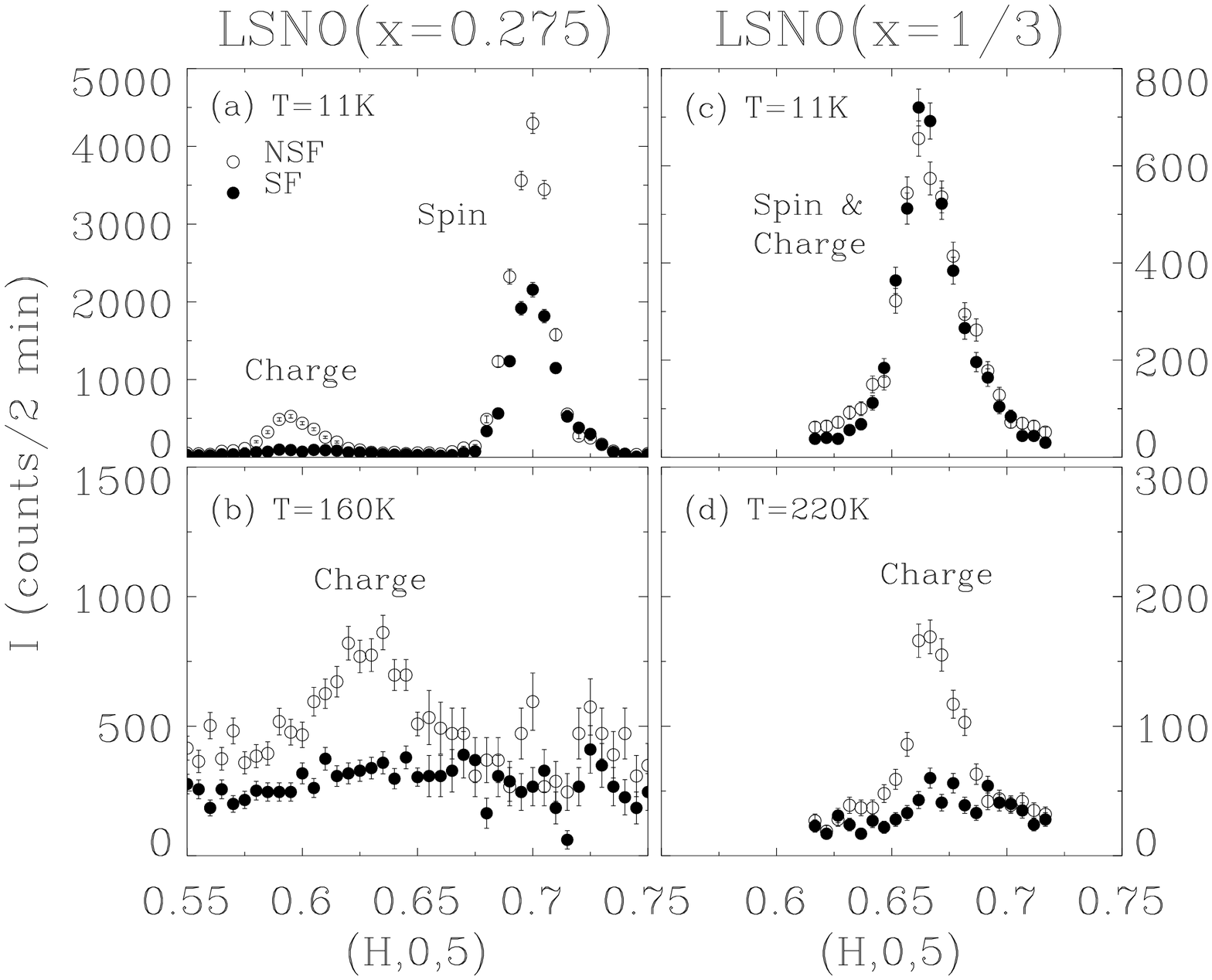,angle=90,width=3.2in}
{Fig.~1. \small
Polarized elastic neutron scattering data along (h05) 
from LSNO(x=0.275) ((a) and (b)), and from LSNO(x=1/3) ((c) and (d)).}
\label{fig:h05}}
\vspace{0.05in}

Fig. 1 shows the results of polarized neutron diffraction
on LSNO(x=0.275 and 1/3). For x=0.275,
at 11K the data exhibit two types of superlattice peaks:
the first harmonic at (0.7,0,5) separated by $(\epsilon,0,0)$
from the (105) AFM
Bragg reflection in pure $\rm La_2NiO_4$
and the second harmonic at (0.6,0,5) 
separated by $(2\epsilon,0,1)$
from the (004) nuclear Bragg reflection
with $\epsilon = 0.3 \approx x$\cite{yoshizawa}.
The first harmonic has
both NSF and SF components whereas the second harmonic
has a NSF component only. These provide 
direct and unambiguous evidence
that the first harmonic is magnetic (spin peak)
whereas the second harmonic is non-magnetic (charge peak) in origin,
and suggest the formation of charge stripes acting as AFM domain
walls (see Fig. 2). 
As shown in Fig. 3 (a), the charge peak gradually increases
below $T_c$ and then get enhanced in a weakly first order
at $T_s$ when spins order.
Inplane charge correlation length, $\xi_c$, also increases at $T_s$.
In the charge and spin ordered phase the charge peak is broader
than the spin peak, indicating that $\xi_c$ is 
shorter than inplane spin correlation length, $\xi_s$:
$\xi_c/\xi_s\approx 100$ \AA/300 \AA~$= 1/3$ (see the inset of Fig. 3 (a)). 
It has been argued\cite{zachar} that disorders in the stripes are mostly due
to non-topological elastic deformations along the stripes
and the decrease of the correlation lengths is inversely proportional
to a power of the periodicity, which can explain why
$\xi_c$ is smaller than $\xi_s$ even though the charge and spin
correlations are inexorably connected.
For comparison, $\xi_c/\xi_s=1/4$
for $\rm La_{1.6-x}Nd_{0.4}Sr_xCuO_4$\cite{tranq99}.
As shown in Fig. 1 (a) and (b), upon heating the (0.7,0,5)
and (0.6,0,5) superlattice peaks
shift toward $(\frac{2}{3},0,5)$ until the (0.7,0,5) peak vanishes 
at 155(5) K and the (0.6,0,5) peak  at 200(10) K.
In the intermediate temperature range $(155 K < T < 200 K)$, as shown in
Fig. 1 (b), the charge order exists without the spin order.
For x=1/3, at 11 K the first and second harmonics with $\epsilon=x$
come together at the same
wave vector as shown in Fig. 1 (c). Unlike the case for x=0.275,
the position of the superlattice peak for x=1/3 is independent
of $T$, implying the stability of commensurability\cite{shl97}. 
At 220 K, the (2/3,0,5) peak has a NSF component
only, indicating that the x=1/3 system also has an 
intermediate $T$ phase with charge order only.

\vspace{0.15in}
\noindent
\parbox[b]{3.4in}{
\psfig{file=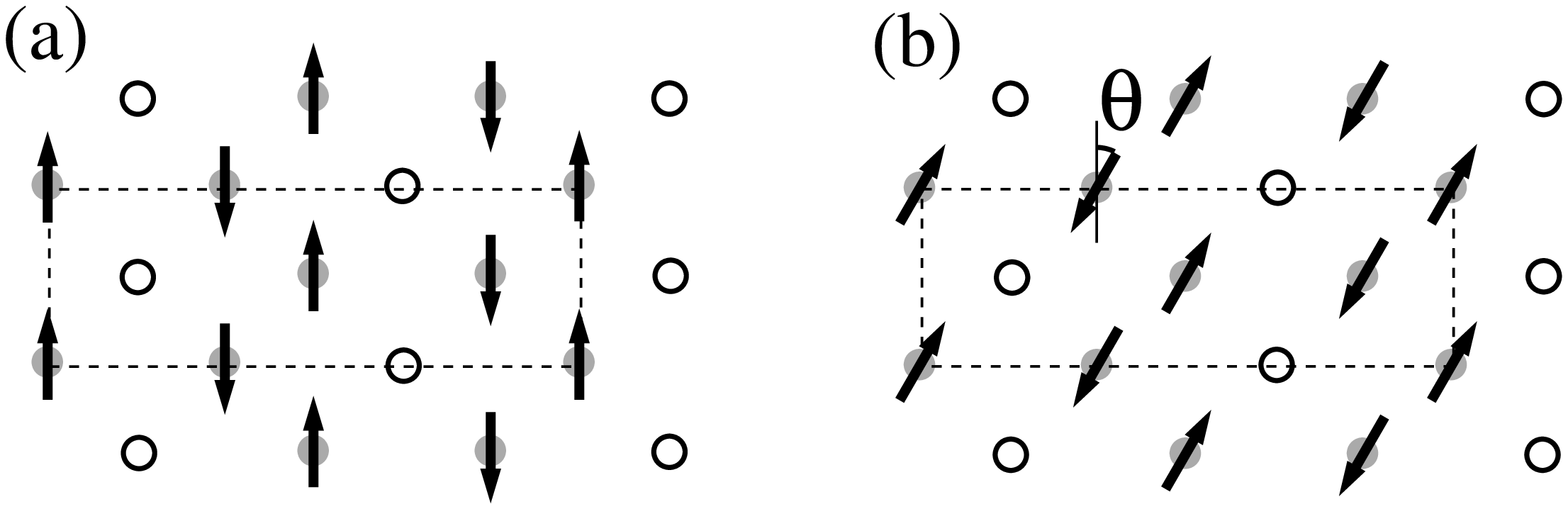,angle=90,width=3.2in}
{Fig.~2. \small
Possible spin structures in a NiO$_2$ plane for LSNO(x=1/3). 
(a) Spins are perpendicular to
the propagation vector along the (100)-axis.
(b) Spins are uniformly
rotated by an angle $\theta$ from the (010)-axis.
Dashed lines
indicate the magnetic unit cell.
}
\label{fig:spinstruct}}
\vspace{0.05in}

Besides identifying the origins of the scattering,
polarized neutron diffraction data also allow us to determine
the spin structure in the material.
Firstly, it is to be noted that the spin peaks 
in the spin ordered states of both samples
are SF in nature. If spins are aligned along the (010)-direction 
(see Fig. 2 (a)), 
$\sigma_M^a=0$ and Eq. (1) would yield zero SF scattering.
There are two possible scenarios to produce nonzero SF scattering.
One possibility is two magnetic domains with a propagation
vector along (100)-axis: in one domain spins are
perpendicular to  and in the other parallel to the propagation vector. 
The population of the domains
must be unequal to explain our data, which seems unlikely.
The other possibility is one single magnetic domain
with $\sigma_M^a\neq 0$.

For quantitative studies of the spin structure,
we have studied the $T$-dependence of the NSF and SF scattering
at various superlattice reflections from both samples.
Fig. 3 (b) shows some of the results for x=1/3.
Upon cooling at 240 K NSF scattering at the (2/3,0,5) reflection
develops without SF scattering, indicating charge ordering.
Below around 200 K, SF as well as NSF scattering increases
due to spin ordering in the usual second order fashion.
At 50 K, however, the trend changes: the NSF scattering decreases
whereas the SF increases.
This behavior can be more clearly seen in the ratio of
SF to NSF scattering shown in Fig. 3 (c).
The sharp increase in the ratios below 50K indicates that the phase
transition at 50 K involves reorientation of spins.
For x=0.275, the SF to NSF ratios of the spin peak in the spin-ordered
phase
were smaller than those for the corresponding reflections in the x=1/3 system
and no spin reorientation transition was observed down to 10K.

\vspace{-0.4in}
\noindent
\parbox[b]{3.4in}{
\psfig{file=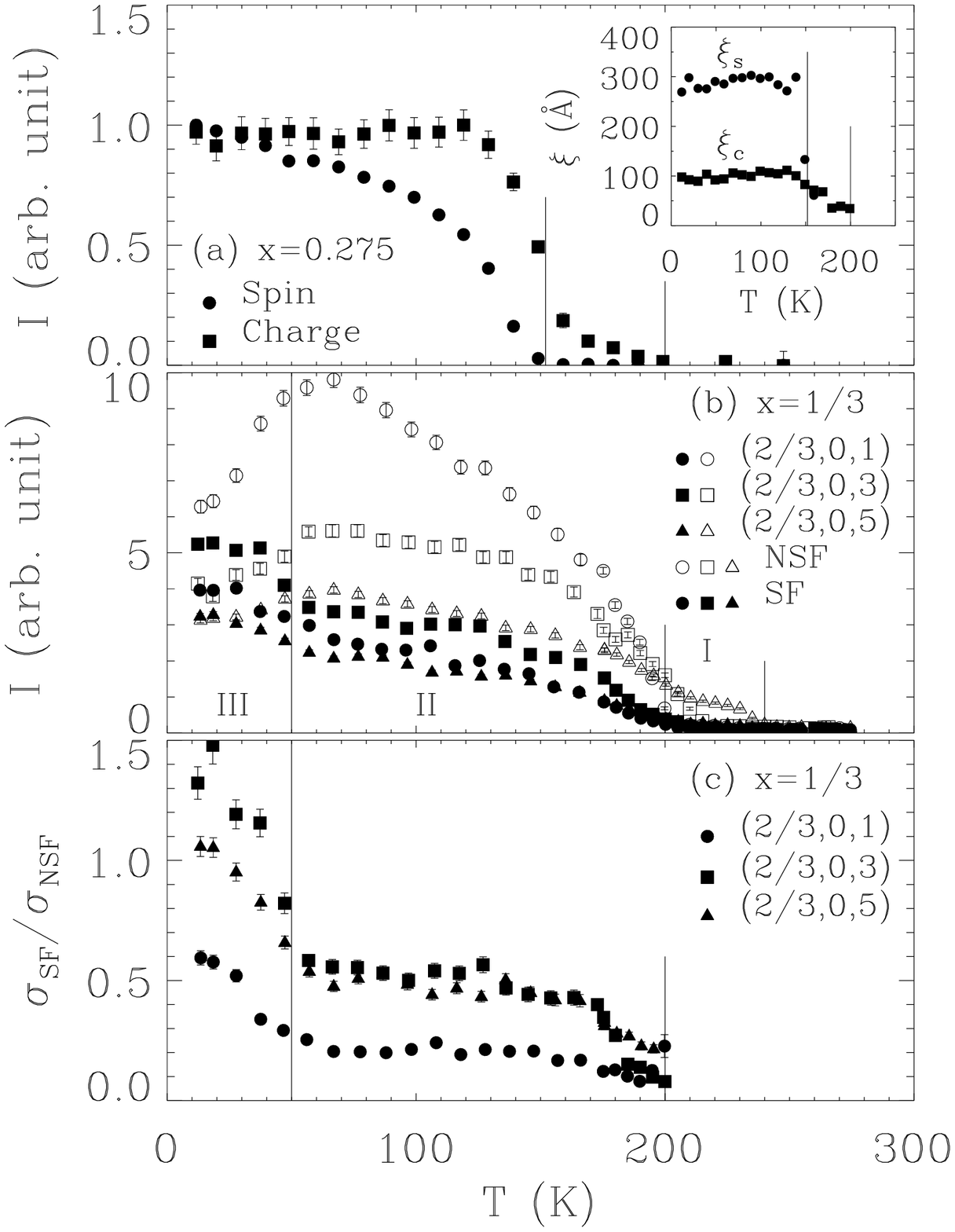,width=2.5in}
{Fig.~3. \small
(a) T-dependence of unpolarized elastic neutron scattering intensities
of the charge peak and the spin
peak. (b) NSF (open symbols) and SF (filled symbols) scattering
intensities at various superlattice reflections from LSNO(x=1/3) as a function
of T. (c) T-dependence of the ratio of SF to NSF scattering intensities
after backgrounds were determined above the transition temperature
and subtracted. Correction for finite polarizing efficiency
was also made\cite{pol_cor}.}
\label{fig:order}}

The symbols in Fig. 4 summarize 
the measured $\sigma_{SF}/\sigma_{NSF}$ 
for LSNO(x=0.275) and for the two spin-ordered phases in LSNO(x=1/3).
The data show a general trend:
for a given value of h in each phase, $\sigma_{SF}/\sigma_{NSF}$
increases as $l$ increases with an exception at
the (2/3,0,5) reflection for x=1/3.
The deviation of the (2/3,0,5) reflection is due to the fact
that the (2/3,0,5) peak has a charge component
as well as spin component. Our unpolarized neutron scattering
data along (0.6,0,$0\leq l \leq 5)$ on the x=0.275 sample 
have the strongest intensity at $l=5$ but negligible at other $l$,
indicating the structure factor due to charge ordering
is strong at $(0.6,0,5)$ but not at other $l$. We expect
the same holds for x=1/3. 

Now consider the collinear spin structure shown in Fig. 2 (b)
in which spins
are rotated by $\theta$ about the c-axis.
Since the spins have a b-component, 
SF scattering would be non-zero and the ratio of SF to NSF scattering
would be determined by the angle $\theta$. 
The lines in Fig. 4 are the calculated
$\sigma_{SF}/\sigma_{NSF}$ for the spin structure shown in Fig. 2 (b):
\begin{equation}
\frac{\sigma_{SF}}{\sigma_{NSF}}=
	{\rm tan}^2\theta \cdot \left( 1-\frac{(ha^*)^2}{(ha^*)^2+(lc^*)^2}\right). 
\end{equation} 
\noindent
\parbox[b]{3.4in}{
\psfig{file=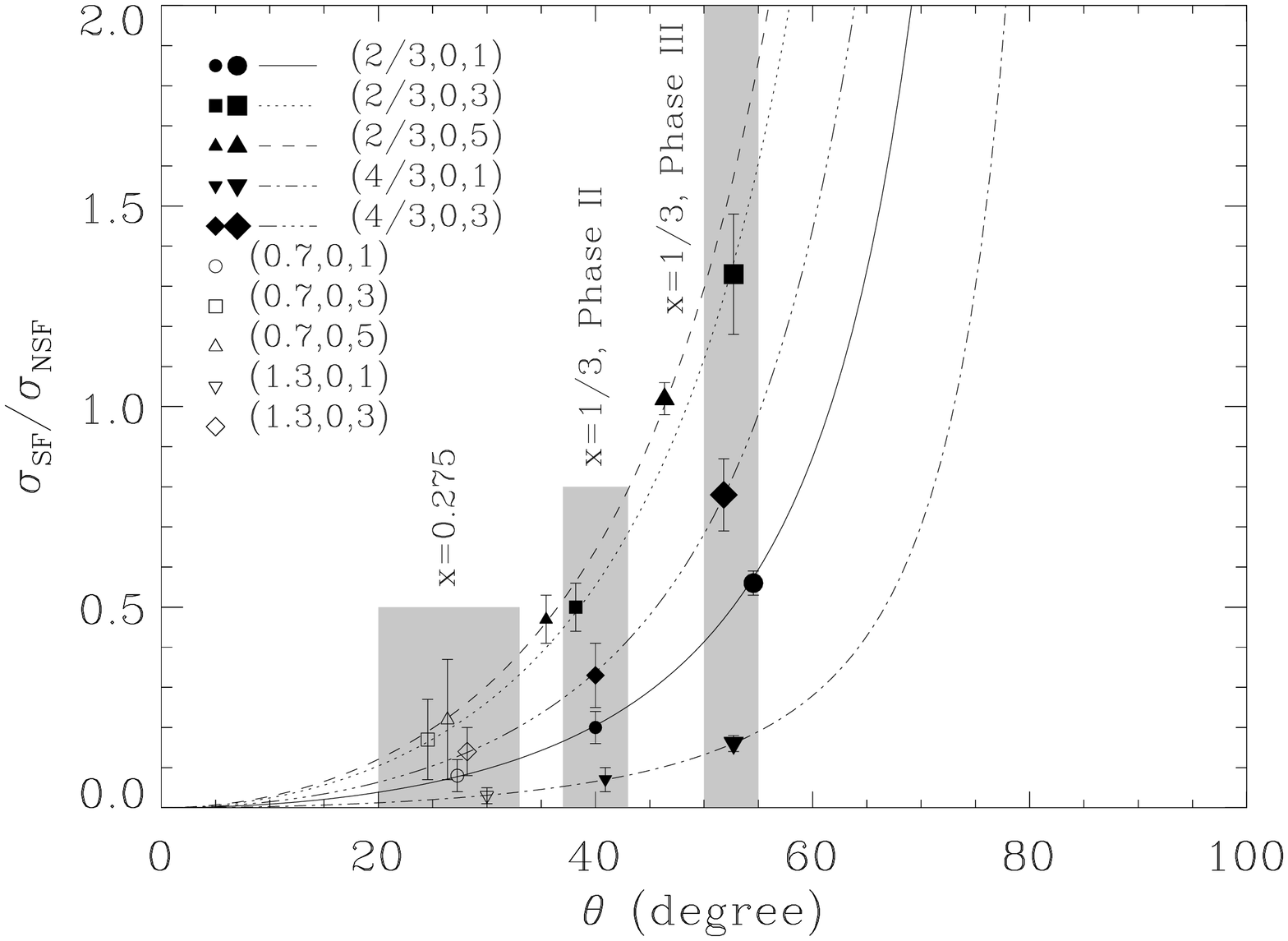,angle=90,width=2.9in}
{Fig.~4. \small
Experimental (symbols) and the model calculated (lines) ratios
of SF to NSF scattering cross sections as a function
of the rotation angle $\theta$.}
\label{fig:model}
}
\vspace{0.05in}

\noindent
From the comparison to the data, we conclude
that $\theta=27^o\pm7^o$ for x=0.275, 
$40^o\pm3^o$ for phase II and $52.5^o\pm2.5^o$ for phase III 
in the x=1/3 sample.
We can also estimate the charge contribution in the (2/3,0,5) reflection
to be $\frac{\sigma_N}{\sigma_M^b}=0.20(6)$.
For x=0.275 (see Fig. 1(a)) 
$\frac{\sigma_N}{\sigma_M^b}=
 \frac{\sigma_{NSF}(0.6,0,5)}{\sigma_{NSF}(0.7,0,5)}=0.12(1)$.

Unpolarized neutron diffraction data along the $l$-direction also contain
information on spin structure.
As shown in Fig. 5, the peaks are much sharper in $l$ 
for x=0.275 than for x=1/3,
which indicates that the correlations in x=0.275 are nearly three-dimensional
whereas those in x=1/3 are quasi-two-dimensional.
Another obvious difference between them is
that the even $l$ to old $l$ peak
intensity ratio, $\frac{I(even ~l)}{I(odd~ l)}$,
is much larger for x=0.275 than for x=1/3.
There are other subtle features to be noted.
For x=0.275, the odd $l$ peak weakens as $l$ increases
whereas the even $l$ peak is strongest at $l=2$.
For x=1/3, at 180K (phase II) the odd $l$ peak also weakens
as $l$ increases. However at 15K (phase III) the $l=3$ peak becomes strongest.
If the spin structure of Fig. 2 (a)  
is displaced by $(\frac{a}{2}n,\frac{1}{2},\frac{1}{2})$
in neighboring NiO$_2$ planes, the magnetic neutron scattering cross section
would become $\sigma (Q)\propto 
|F(Q)|^2(1+(-)^l {\rm cos} n\pi h)$\cite{wochner}
where $F$ is the magnetic form factor of the Ni$^{2+}$.
Therefore, for a given $h$, the $l$-dependence of odd or even $l$ peaks
would just follow $|F(Q)|^2$. This $\sigma (Q)$ cannot
explain those subtle features.
On the other hand, the spin structure of 
Fig. 2 (b) would give
\begin{eqnarray}
\sigma (Q) & \propto & |F(Q)|^2(1+(-)^l {\rm cos} n\pi h) \nonumber \\
 & & \times \left( 1-{\rm sin}^2\theta\frac{(ha^*)^2}{(ha^*)^2+(lc^*)^2}\right).
\end{eqnarray}
With
the $\theta$'s which are obtained from the polarization analysis,
Eq. 3
reproduces the $l$-dependence remarkably well for the case of x=0.275 with
three-dimensional magnetic correlations, as shown as shaded bars
in Fig. 5.

\vspace{-.4in}
\noindent
\parbox[b]{3.4in}{
\psfig{file=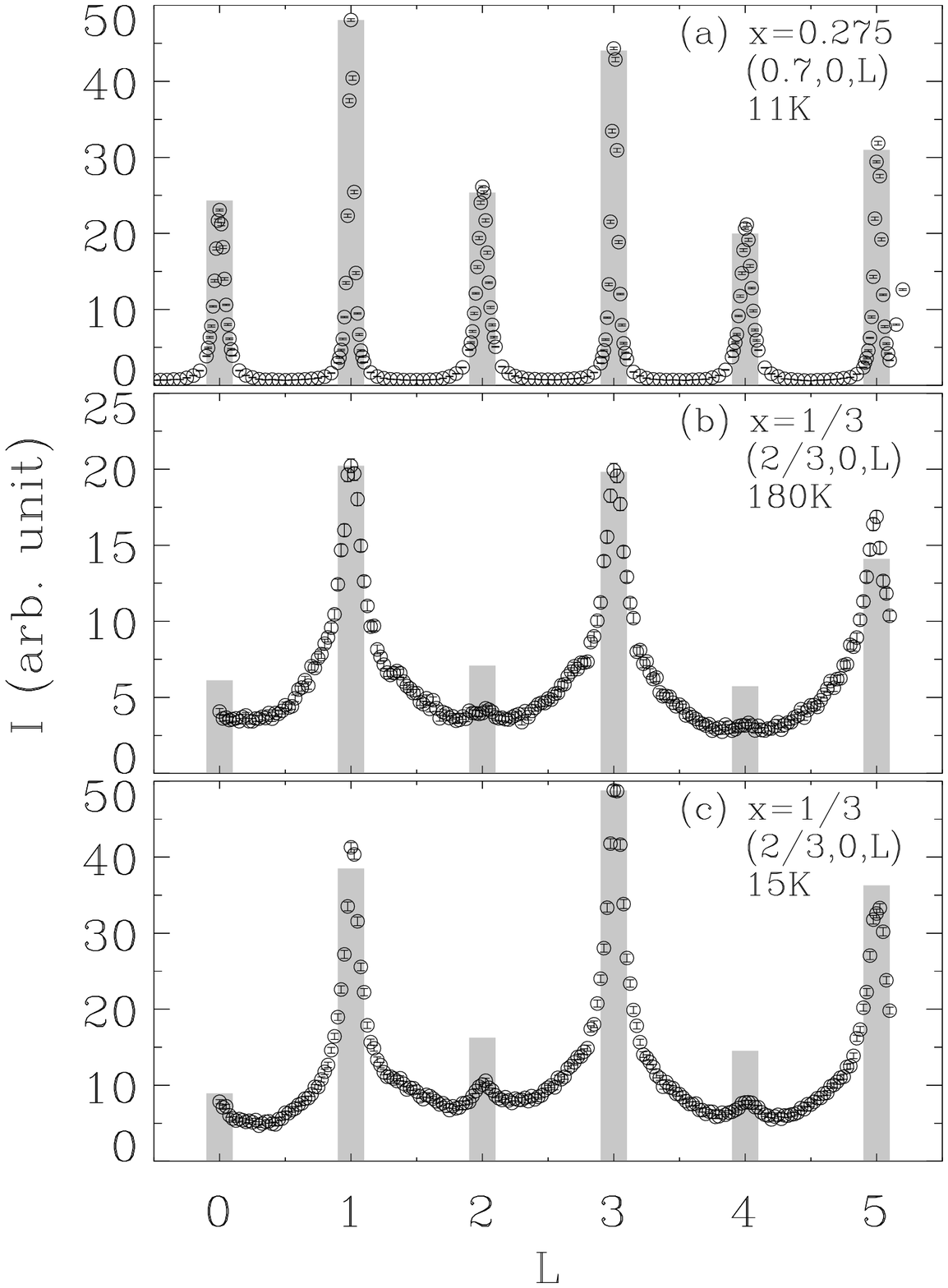,width=2.5in}
{Fig.~5. \small
$l$-dependence of (a) (0.7,0,$l$) scan form LSNO(x=0.275) at 11 K, and
(2/3,0,$l$) scan from LSNO(x=1/3) (b) at 180 K and (c) 15 K.
The shaded bars are obtained by the model which is described in the text.}
\label{fig:ldep}
}
\vspace{0.05in}

\noindent
In the case of x=1/3 where the magnetic order is quasi-two dimensional,
the long tails of the peaks make it difficult to extract the integrated
intensities to compare with the model.
Nevertheless, the model explains the subtle features
such as the (2/3,0,3) peak being considerably stronger 
than the other reflections in phase III.

Spin reorientations upon cooling have been observed in
some isostructural materials such as pure $\rm La_2CoO_4$\cite{yamada89}
and $\rm Nd_2CuO_4$\cite{jeff}.
Those reorientations involve 
spin flip (by $\pi/2$ or $\pi$) which are due to  
either a structural phase transition (in $\rm La_2CoO_4$)
or strong coupling between the rare-earth moments
and $Cu^{2+}$ moments (in $\rm Nd_2CuO_4$).
For LSNO(x=1/3), however, there is 
no structural phase transition 
down to 10 K within our experimental uncertainty\cite{struct}.
What causes then the reorientation of spins in LSNO(x=1/3)
below 50 K ?  
The answer might be a further localization of the holes
on the lattice and their interactions with the surrounding S$=$1
$Ni^{2+}$ moments. Among the two charge ordered states, below and above 50 K,
the holes should be more strongly pinned on the lattice
in the lower $T$ phase. 
The further localization of the holes might either induce
a subtle and yet undetected local crystal distortion around 
the surrounding S$=$1 spins
or magnetically order themselves,
either of which would in turn reorient the S$=$1 spins.
This idea is consistent
with recent resistivity measurements\cite{tokura}.
In the resistivity measurements, a moderate inplane electric field 
($V_{cir}\geq 100$ V)
decreases the resistivity by up to 5 orders of magnitude in phase II.
However, although the charge-ordered state in phase III 
becomes suppressed by the electric field, phase III
survives until $V_{cir} \geq 900$ V, which
indicates the stronger localization of the holes in phase III
than in phase II. 
Understanding the microscopic origin of the further localization in LSNO(x=1/3)
and its effects, such as the spin
reorientation, as well as the evolution of spin structure upon doping
in LSNO(x)
requires 
more theoretical and experimental studies.

The authors thank G. Aeppli, J.M. Tranquada, G. Shirane, 
Y.S. Lee, and S. Wakimoto for helpful discussions. 
Work at SPINS is based upon activities supported by the National
Science Foundation under Agreement No. DMR-9423101.


\begin{references}
\bibitem{cheong91} S-W. Cheong {\it et al.}, 
Phys. Rev. Lett. {\bf 67}, 1791 (1991).
\bibitem{yamada98} 
K. Yamada {\it et al.}, Phys. Rev. B {\bf 57}, 6165 (1998).
\bibitem{tranq95} J.M. Tranquada {\it et al.}, 
Nature {\bf 375}, 561 (1995); Phys. Rev. Lett. {\bf 78}, 338 (1997).
\bibitem{emery99} 
V. J. Emery, S. A. Kivelson, and J. M. Tranquada, Proc. Natl.
Acad. Sci. {\bf 96}, 8814 (1999) and references therein.
\bibitem{aeppli88} G. Aeppli and D.J. Buttrey, Phys. Rev. Lett. {\bf 61},
203 (1988).
\bibitem{hayden92} S.M. Hayden {\it et al.}, Phys. Rev. Lett. {\bf 68}, 1061
(1992). 
\bibitem{chen93} C.H. Chen, S-W. Cheong and A.S. Cooper,
Phys. Rev. Lett. {\bf 71}, 2461 (1993).
\bibitem{tranq94} J.M. Tranquada {\it et al.}, 
Phys. Rev. Lett. {\bf 73}, 1003 (1994); 
Phys. Rev. B {\bf 52}, 3581 (1995); 
Phys. Rev. B {\bf 54}, 12318 (1996).
\bibitem{yoshizawa} H. Yoshizawa, {\it et al.}, cond-mat/9904357.
\bibitem{tokura} S. Yamanouchi {\it et al.},
Phys. Rev. Lett. {\bf 83}, 5555 (1999).
\bibitem{hk0} Our polarized neutron diffraction measurements
on the LSNO(x=1/3) sample in the (hk0) scattering plane 
confirmed that the spins lie in the ab-plane.
\bibitem{moon} R.M. Moon {\it et al.}, Phys. Rev. {\bf 181},
920 (1969). 
\bibitem{zachar} O. Zachar, cond-mat/9911171.  
\bibitem{tranq99} J.M. Tranquada {\it et al.}, 
Phys. Rev. B {\bf 59}, 14712 (1999). 
\bibitem{shl97} S.-H. Lee and S-W. Cheong,
Phys. Rev. Lett. {\bf 79}, 2514 (1997).
\bibitem{pol_cor} C.F. Majkrzak, Physica B {\bf 221}, 342 (1996).
\bibitem{wochner} P. Wochner {\it et al.}, 
Phys. Rev. B {\bf 57}, 1066 (1998).
\bibitem{yamada89} 
K. Yamada {\it et al.}, Phys. Rev. B {\bf 39}, 2336 (1989). 
\bibitem{jeff} 
S. Skanthakumar {\it et al.}, Physica C {\bf 160}, 124 (1989).
\bibitem{struct} 
Neither a splitting of the lattice constant $a$ 
(orthorhombic distortion)
nor a superlattice reflection at (102) (allowed for
another tetragonal P4$_2$/ncm structure) has been found
in LSNO(x=1/3) for 10 K $<T<$ 320 K.






\end{references}
\end{document}